\RequirePackage{fix-cm}
\documentclass[twocolumn,epjc3]{svjour3}  
\smartqed  
\RequirePackage{graphicx}
\usepackage{comment}
\usepackage{lineno}
\usepackage{csquotes}
\usepackage{cite}
\usepackage{orcidlink}
\journalname{Eur. Phys. J. C}
\begin{document}

\title{Study of proton-nucleus interactions in  the DSTAU/NA65 experiment at the CERN-SPS
}


\author{S. Aoki\orcidlink{0000-0002-1092-5037}\thanksref{addr1},
        A. Ariga\orcidlink{0000-0002-6832-2466}\thanksref{addr2,addr3}, T. Ariga\orcidlink{0000-0001-9880-3562}\thanksref{addr4}, N. Charitonidis\orcidlink{0000-0001-9506-1022}\thanksref{addr5}, S. Dmitrievsky\orcidlink{0000-0003-4247-8697}\thanksref{addr6},  R. Dobre\orcidlink{0000-0002-9518-6068}\thanksref{addr7}, 
        N. Ergin\orcidlink{0009-0001-6196-7974}\thanksref{addr8}, 
        E. Firu\orcidlink{0000-0002-3109-5378}\thanksref{addr7}, Y. Gornushkin\orcidlink{0000-0003-3524-4032}\thanksref{addr6}, A. M. Guler\orcidlink{0000-0001-5692-2694}\thanksref{a,addr5,addr8}, D. Hayakawa\orcidlink{0000-0003-4253-4484}\thanksref{addr2}, 
        F. Kling\orcidlink{0000-0002-3100-6144}\thanksref{addr13},
        K. Kodama\orcidlink{0000-0001-9533-1571}\thanksref{addr9}, M. Komatsu\orcidlink{0000-0002-6423-707X}\thanksref{addr10}, U. Kose\orcidlink{0000-0001-5380-9354}\thanksref{addr11},  M. Miura\orcidlink{0000-0002-4955-8609}\thanksref{addr2}, T. Nakano\orcidlink{0009-0004-8568-9077}\thanksref{addr10}, A. T. Neagu\orcidlink{0000-0001-6788-4320}\thanksref{addr7}, T. Okumura\orcidlink{0000-0002-3266-8713}\thanksref{addr2}, 
        C. Oz\orcidlink{0000-0002-5113-5779}\thanksref{addr8}, H. Rokujo\orcidlink{0000-0002-3502-493X}\thanksref{addr10}, O.  Sato\orcidlink{0000-0002-6307-7019}\thanksref{addr10}, S. Vasina\orcidlink{0000-0003-2775-5721}\thanksref{addr6}, M. Yoshimoto\orcidlink{0000-0002-4667-0718}\thanksref{addr12}, E. Yuksel\orcidlink{0009-0008-7861-1879}\thanksref{addr8}   
        }
\thankstext{a}{e-mail: ali.murat.guler@cern.ch}
\institute{Graduate School of Human Development and Environment, Kobe University, Tsurukabuto, Nada, 657-8501 Kobe, Japan\label{addr1}\and
           Chiba University, Yayoi 1-33, Inage, Chiba, 263-8522 Chiba, Japan\label{addr2}\and
           Albert Einstein Center for Fundamental Physics, Laboratory for High Energy Physics, University of Bern, Sidlerstrasse 5, CH-3012 Bern, Switzerland\label{addr3}\and
           Kyushu University, 744 Motooka, Nishi-ku, Fukuoka, 819-0395 Japan \label{addr4}\and
           CERN, 1 Esplanade des Particules, CH-1211 Meyrin, Switzerland \label{addr5}\and 
           Affiliated with an international laboratory covered by a cooperation agreement with CERN \label{addr6}\and 
           Laboratory of High Energy Physics, Institute of Space Science subsidiary of INFLPR, 409, Atomistilor Street, Magurele, 077125 Ilfov, Romania \label{addr7}\and
            Deutsches Elektronen-Synchrotron DESY, Notkestr.~85, 22607 Hamburg, Germany\label{addr13}
           \and
           Physics Department, Middle East Technical University, Dumlup{\i}nar Bulvar{\i}, 06800 Ankara, T{\"u}rkiye \label{addr8}\and 
           Department of Science Education, Aichi University of Education, 448-8542 Kariya, Japan \label{addr9}\and
           Nagoya University, Furo-cho, Chikusa-ku, 464-8602 Nagoya, Japan \label{addr10}\and 
           Institute for Particle physics and Astrophysics, ETH Zurich, Otto-Stern-Weg 5, CH-8093 Zurich, Switzerland \label{addr11}\and
           RIKEN Nishina Center, RIKEN, 2-1 Hirosawa, Wako, 351-0198 Saitama, Japan \label{addr12}
}

\date{Received: date / Accepted: date}
\maketitle
\begin{abstract}

The DsTau(NA65) experiment at CERN was proposed to measure an inclusive differential cross-section of $D_s$ production with decay to tau lepton and tau neutrino in $p$-$A$ interactions. 
The DsTau detector is based on the nuclear emulsion technique, which provides excellent spatial resolution for detecting  short-lived particles like charmed hadrons. This paper presents the first results of the analysis of the pilot-run (2018 run) data and reports the accuracy of the proton interaction vertex reconstruction.
High precision in vertex reconstruction enables detailed measurement of proton interactions, even in environments with high track density. 
The measured data has been compared with several Monte Carlo event generators in terms of multiplicity and angular distribution of charged particles. The multiplicity distribution obtained in p-W interactions is tested for KNO-G scaling and is found to be nearly consistent.
The interaction length of protons in tungsten is measured to be 93.7 $\pm$ 2.6 mm. 
The results  presented in this study can be used to validate   event generators of $p$-$A$ interactions.
\end{abstract}

\section{Introduction}
\label{intro}
The DsTau experiment \cite{dstau} aims to measure tau neutrino production in $p$-$A$ interactions at  the CERN-SPS.
The study  of tau neutrino interactions is an important probe in constraining the
 models beyond the Standard Model. For example,  the lepton flavour universality of the weak interaction can  be tested in the context of neutrinos as complementary to measurements in the LHCb \cite{lhcb}  and  Belle II \cite{belle2} experiments.  An accurate knowledge of tau neutrino flux in accelerator neutrino beams is 
 essential for ongoing and future neutrino experiments like FASER \cite{faser}, SND@LHC \cite{snd}, SHiP  \cite{ship}.
 In the past, only a few experiments  reported tau neutrino interactions with low statistics.  
The first experiment that directly observed the tau neutrino charged-current interactions was  DONuT \cite{donut}  at Fermilab in 2000.
Later, tau neutrino interactions were also detected by  the OPERA \cite{opera}, Super-K \cite{superk}, and IceCube \cite{icecube}  experiments. However, in these experiments, tau neutrino measurements are affected by  neutrino oscillations. The DONuT experiment provided an estimate of the tau neutrino interaction cross-section \cite{donut2}, although this estimate is accompanied by considerable uncertainty. The systematic inaccuracy, which arises from uncertainties in the tau neutrino production mechanism and is approximately 50$\%$ relative to the cross-section value. Furthermore, the statistical error accounts for about 33$\%$, as only nine $\nu_\tau$ events were detected.
DsTau aims to reduce the systematic uncertainty 
in the tau neutrino production down to 10$\%$  level by  detecting about 
$10^3$ $D_s\rightarrow\tau$ decays in $2\times 10^8$ proton interactions  in the tungsten or molybdenum target. Moreover, the charmed hadron pair production in $p$-$A$ interactions can also be studied with  large statistics. 

In DsTau, the data analysis is characterized by a very high track density and  pile-up of events and is performed in two steps. The first step  is track recognition and  primary proton vertex reconstruction. The second one is to search for charm particles by their decay topology.  
In this paper, we present the first results related to the  primary proton interaction analysis based on the 2018 run data sub-sample. The data quality in terms of accuracy and statistics would allow us to perform the analysis
of the primary proton interactions which  
provides  a measurement of the multiplicity  and angular distribution of charged secondary particles in different target materials.  
Those features are compared with several widely used Monte Carlo (MC) generators. Our results can therefore be used to tune the interaction models implemented in the generators. 
\section{Experimental setup and data taking}
\label{sec:1}
\begin{figure}[ht]
  \includegraphics[width=0.45\textwidth]{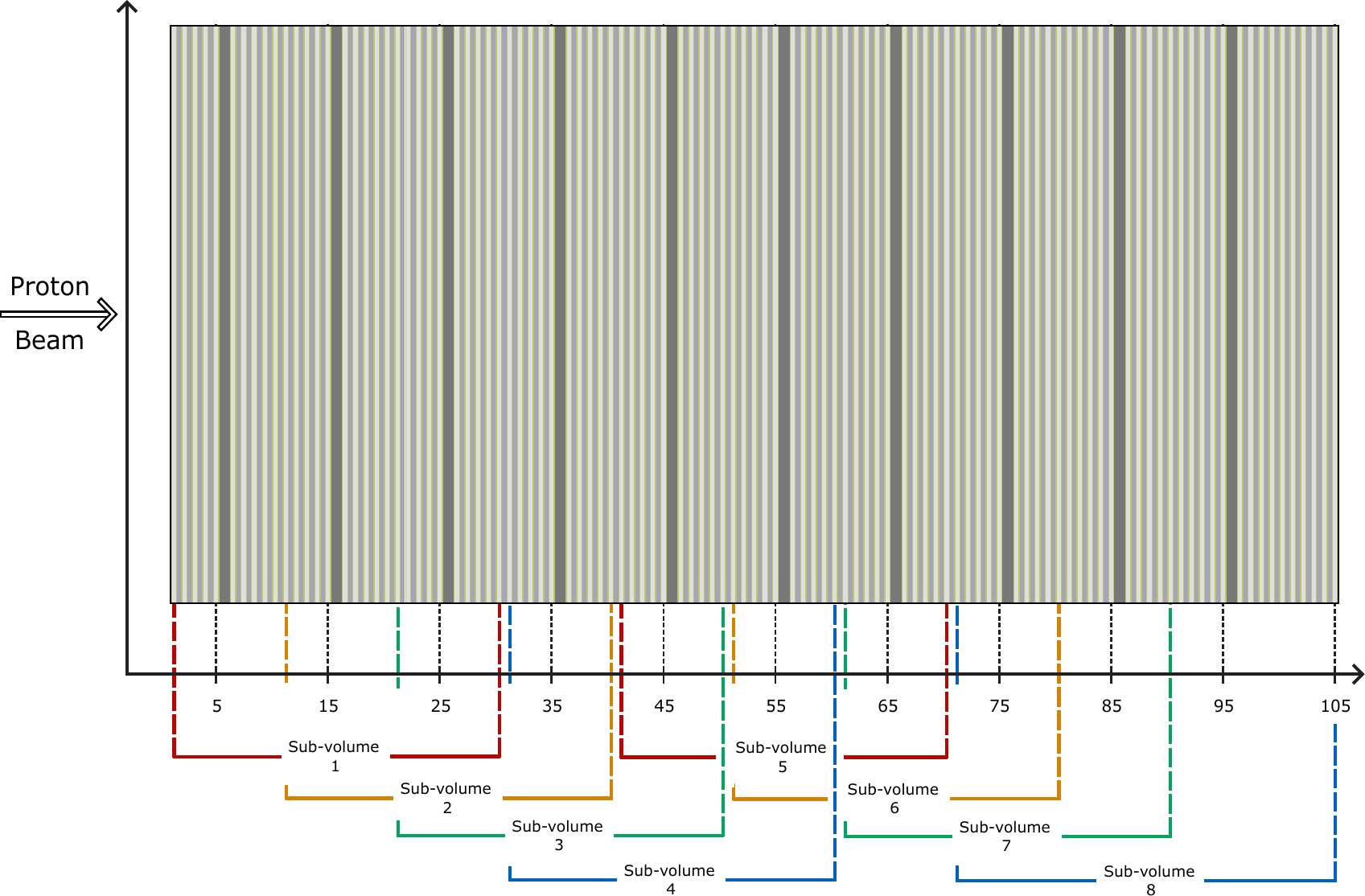}
\caption{Schematic view of a module used in the 2018  run. A module consists of 10 tungsten plates and 105 emulsion plates. A module is divided into 8 sub-volumes, each consisting of three tungsten plates and 30 emulsion films}
\label{fig:1}       
\end{figure}

Observing the decay topology of $D_s$ into $\tau$ requires sub-micron spatial resolution and sub-mrad three-dimensional angular resolution. Among  the available detector technologies, 
only nuclear emulsion\cite{tomoko}  can provide the necessary spatial and angular resolutions.
The DsTau detector, based on the nuclear emulsion technique,  consists of tungsten/molybdenum  plates inter-spaced with nuclear emulsion films and plastic spacers.
Tungsten and molybdenum (used only  in Physics runs 2021-2023) have been chosen as a target as   they are  used for neutrino beam generation at accelerators (i.e. in case of DONuT and SHiP experiments). 
In a DsTau module, the tungsten or molybdenum plates  act as a target for beam protons; emulsion films separated by plastic spacers that act as a decay volume and a high-accuracy tracking device. 
The nuclear emulsion films consist of 70-$\mu$m thick  emulsion layers on both sides of a 210--$\mu$m thick plastic base.
A schematic view of one module and one unit  used in the 2018 run is shown in  Fig.~\ref{fig:1}.
In each unit, there is a 500 $\mu$m tungsten plate followed by 10 nuclear emulsion films interleaved with 9 plastic spacers. This unit structure is repeated 10 times in a module that is 12.5 cm wide, 10 cm high, and 4.8 mm thick.
Additionally, there are five emulsion films with plastic spacers placed upstream to tag beam protons. The purity of tungsten plates is greater than 99.95$\%$, as measured by Goodfellow Cambridge Ltd.
An Emulsion Cloud Chamber, comprising 26 emulsion films interleaved with 1 mm thick lead plates, is placed downstream of the module to reconstruct secondary particle momentum by measuring multiple Coulomb scattering.  
Modern use of nuclear emulsion is based on impressive progress in the scanning technique achieved during the last two decades thanks to the pioneering works by Nagoya University group \cite{aoki,nakano}   and the further development in the frame of the OPERA experiment \cite{opera}.
At  present, the most advanced 
scanning system called  the Hyper Track
Selector,  HTS \cite{hts}, reads out track information from two  layers 
 of the emulsion film at an average speed of 1000 cm$^2$/h at the DsTau.
The HTS captures 22 tomographic images at equally spaced depths of the emulsion layer. These images are then digitized, and an image processor produces binary images that separate silver grains recorded by a penetrating track from the background.
 The online scanning program provides the positions and angles of the detected micro-tracks, which are a chain of aligned clusters in one emulsion layer. The base-tracks are obtained by linking two micro-tracks within the angle and position tolerances across the plastic base.
The base-track reconstruction efficiency is measured to be greater than 90$\%$ for tracks with an angle of less than 0.4 radians with respect to the direction perpendicular to the emulsion surface. The efficiency gradually decreases by a few percentage points towards the downstream part of the detector as the track density increases. However, by allowing the absence of one or more  base tracks in the track reconstruction, the overall efficiency for reconstructing tracks remains sufficiently high, more than  99$\%$.

The CERN Super Proton Synchrotron  provides a beam of protons with a momentum of 400 GeV/c.  In 2016 and 2017, test beam studies were carried out to evaluate and characterize the detector concept. In 2018, a pilot run was conducted to demonstrate the detection of proton interactions in a high track density environment and to record 10$\%$  of the planned experimental data. During the pilot run, 30 modules were exposed to the proton beam at the H4 beam line. Each emulsion module was mounted on a motorized X-Y stage, called the target mover\cite{tora}, which allowed synchronized movement of the module with respect to the proton spills. This setup ensured uniform irradiation of the detector surface with a density of $10^5$ tracks/$\mathrm{cm^2}$.
The emulsion scanning of the 2018 run films has been completed, and  the analysis of the collected
data is ongoing. For the present measurement, we report the analysis results of a sub-data sample   from the 2018 run~\cite{dstau}. 

\section{Analysis}
\label{sec:2}
\begin{figure}[ht]
  \includegraphics[width=0.48\textwidth]{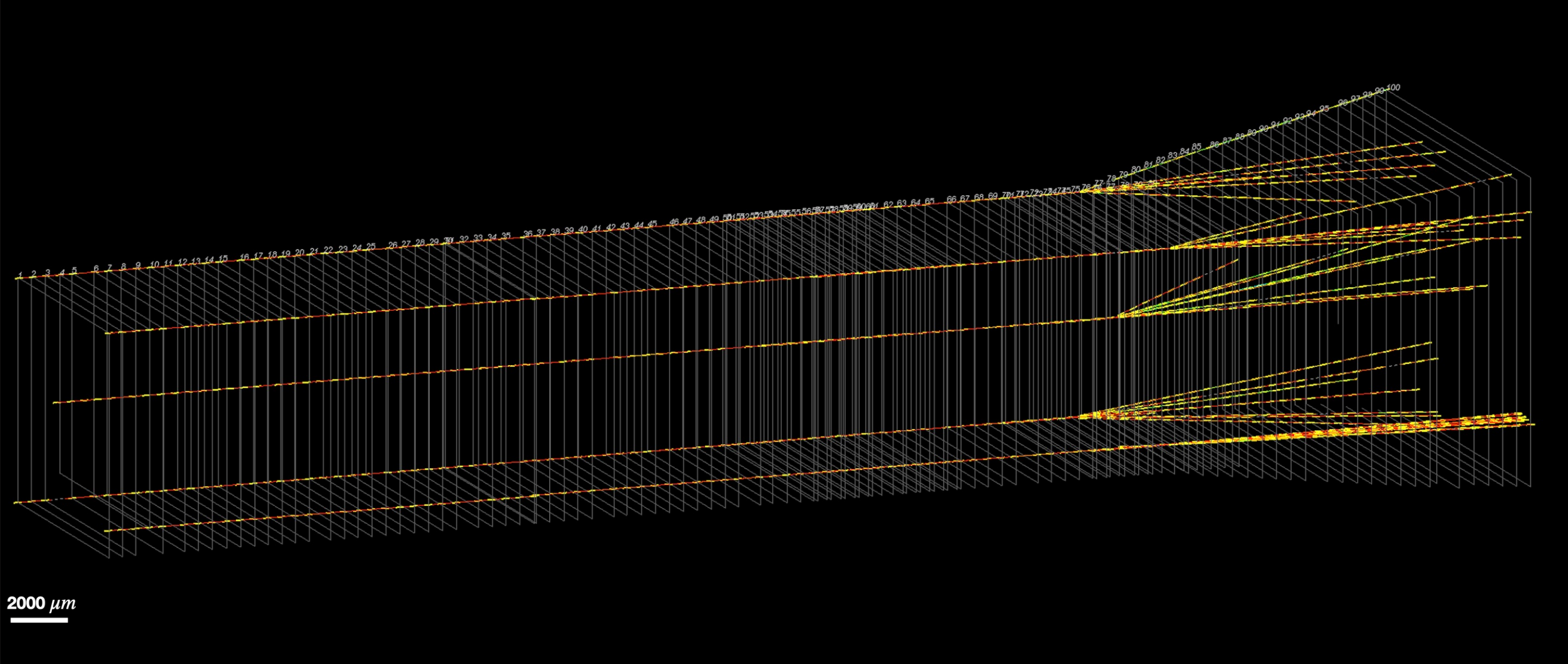}
\caption{The  reconstructed proton interactions in the area of 1 $\mathrm{cm^2}$ in the full module of the 2018 run data (only a few of them are shown for  visibility)}
\label{fig:2}       
\end{figure}

\begin{figure}[ht]
  \includegraphics[width=0.4\textwidth]{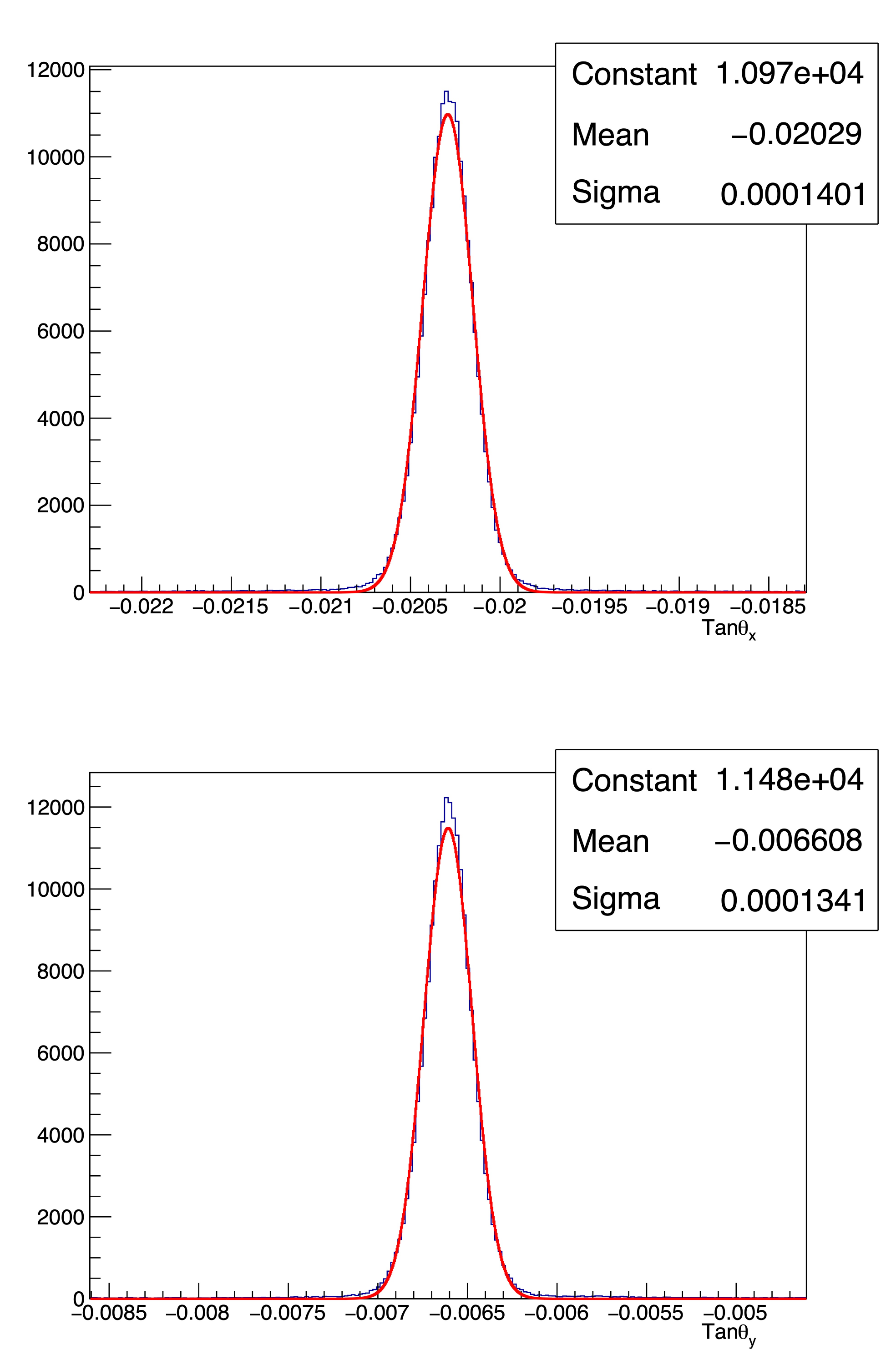}
\caption{The distribution of the proton track angle in the data is shown for the XZ and YZ planes, with a Gaussian fit}
\label{fig:3}       
\end{figure}

After the  scanning   by the HTS, the emulsion data is processed by the DsTau software to reconstruct tracks and vertices. To process the data efficiently, each module is divided into 8 overlapping volumes along the longitudinal direction. Each volume contains three tungsten plates and 30 emulsion films, with the first sub-volume including an additional five emulsion plates(Fig. 2). Furthermore, each volume is divided into 63 sub-volumes along the transverse plane, with each sub-volume having a size of 1.5 cm $\times$ 1.5 cm $\times$ 30-35 emulsion films.
After base-track reconstruction,  film-to-film alignment is performed by applying base-track pattern matching.
To define a global reference system, a set of affine transformations
has to be computed to account for the different reference frames used for data taken
in different emulsion films. Once all emulsion films are aligned, base-tracks are connected film by film according to angular and position acceptance to form  so-called volume tracks. The track finding and fitting consider possible
inefficiencies in the base-track reconstruction. 
Further alignment between films (a scale factor, a rotation, transverse position shifts, and a gap) is obtained by using tracks, reconstructed in each sub-volume. In particular, for the transverse position alignment, the 400-GeV proton beam track, supposed to be most straight, is selectively used. 
After completing the track-reconstruction step, vertexing algorithms are applied to identify the proton interactions. Tracks containing at least four base-tracks are used for vertexing. The algorithm uses the points of the closest approach of the multiple tracks to determine the vertex's position with high accuracy. To ensure high purity in the vertex reconstruction, vertices with more than four tracks are used for further analysis.
The beam protons in the module are selected based on the track angle, obtained by fitting base-tracks in the first 5 emulsion films to a straight line. The proton angular distribution is presented in Fig. ~\ref{fig:3} with a Gaussian fit. If the track angle falls within 4$\sigma$ of the measured mean beam angle, it is identified as a beam proton and subsequently followed down through the module in a process referred to as proton-linking~\cite{emin}. 
The purity of beam proton selection  is estimated  to be  $>96\%$ from the MC simulation. 
Following the linking process, the number of beam protons traversing each tungsten plate is estimated, as shown in Table~\ref{tab:11}.
The beam proton track is subsequently linked to
the reconstructed vertices through a method called proton-vertex linking. A vertex is identified as a proton interaction vertex if its parent matches in both angle and position with one of the beam protons selected in proton-linking step. This procedure eliminates contamination from secondary particle interactions, which progressively increases with depth in the module. Tungsten plates and the other materials in a module generate hadrons through proton interactions and act as radiators to produce electromagnetic showers. Consequently, the track density rises from about $10^5$ to $4.5\times 10^5$ tracks/$\mathrm{cm^2}$ along the longitudinal direction. Despite this high track density, proton interaction vertices can be  fully reconstructed within  the module, with several examples illustrated in Fig.~\ref{fig:2}.
For the present analysis, 
a sample of 95,314 events that have a  reconstructed beam-proton and its interaction vertex in the tungsten plate is  used from  a single module of the 2018 run (Table \ref{tab:11}).
 To ensure high efficiency and purity in track and vertex reconstruction,  the data related to interactions in the last two
 tungsten plates are not included.

\subsection{Efficiency evaluation}

\begin{figure}[ht]
    \centering
        \includegraphics[width=0.45\textwidth]{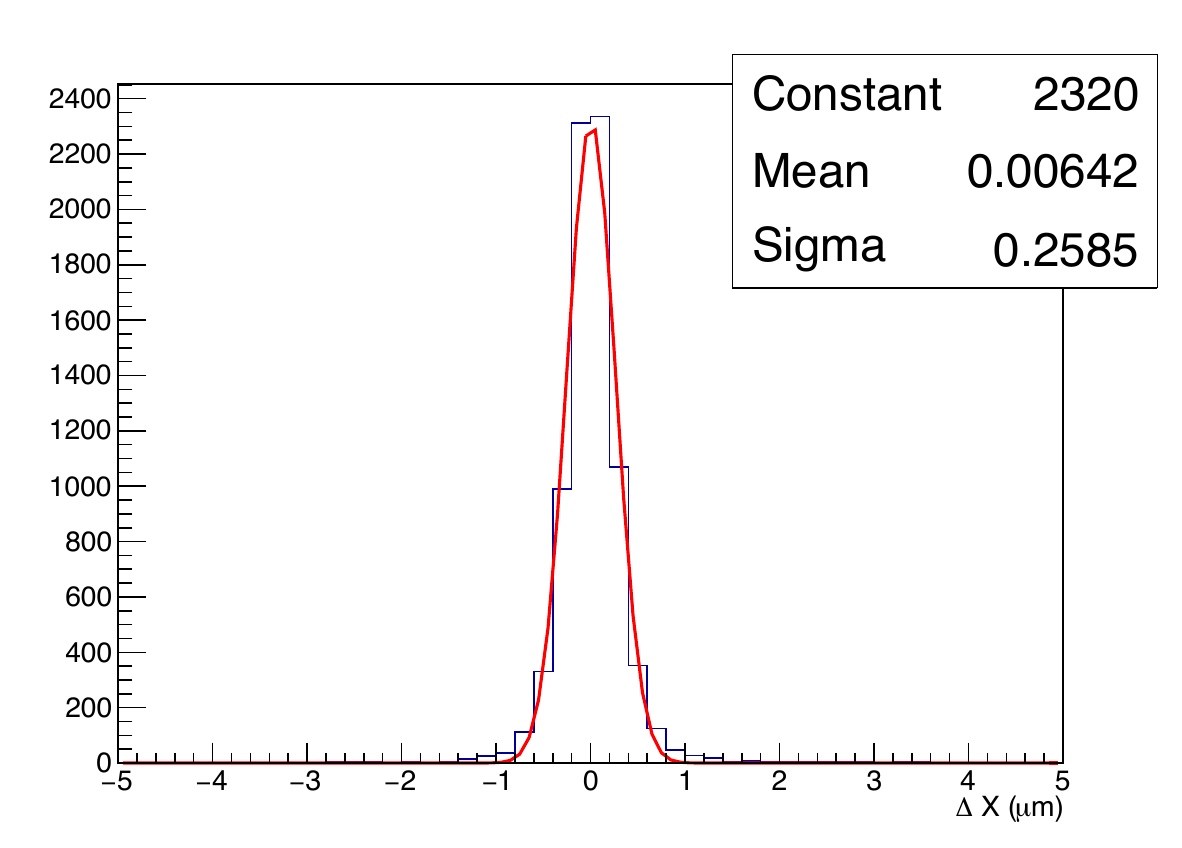}
    \vspace{1em} 
        \includegraphics[width=0.45\textwidth]{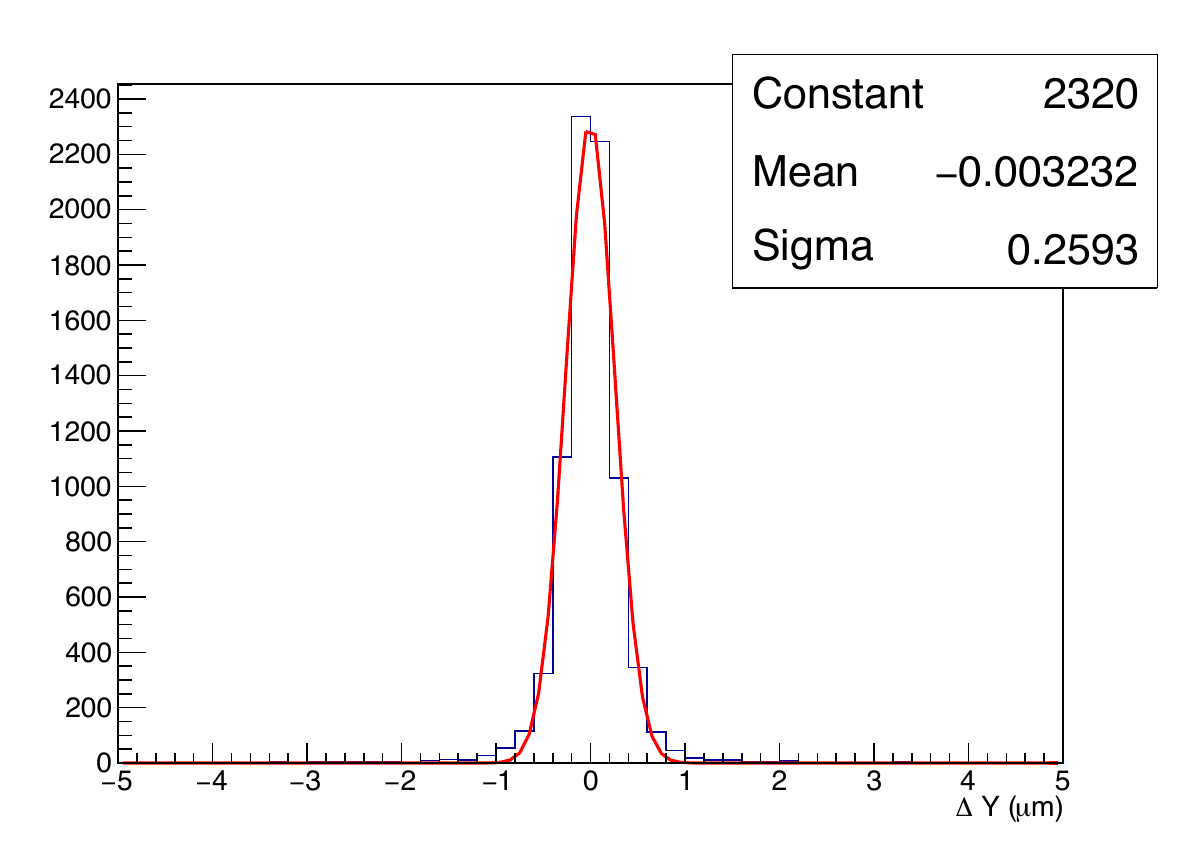}
    \vspace{1em} 
        \includegraphics[width=0.45\textwidth]{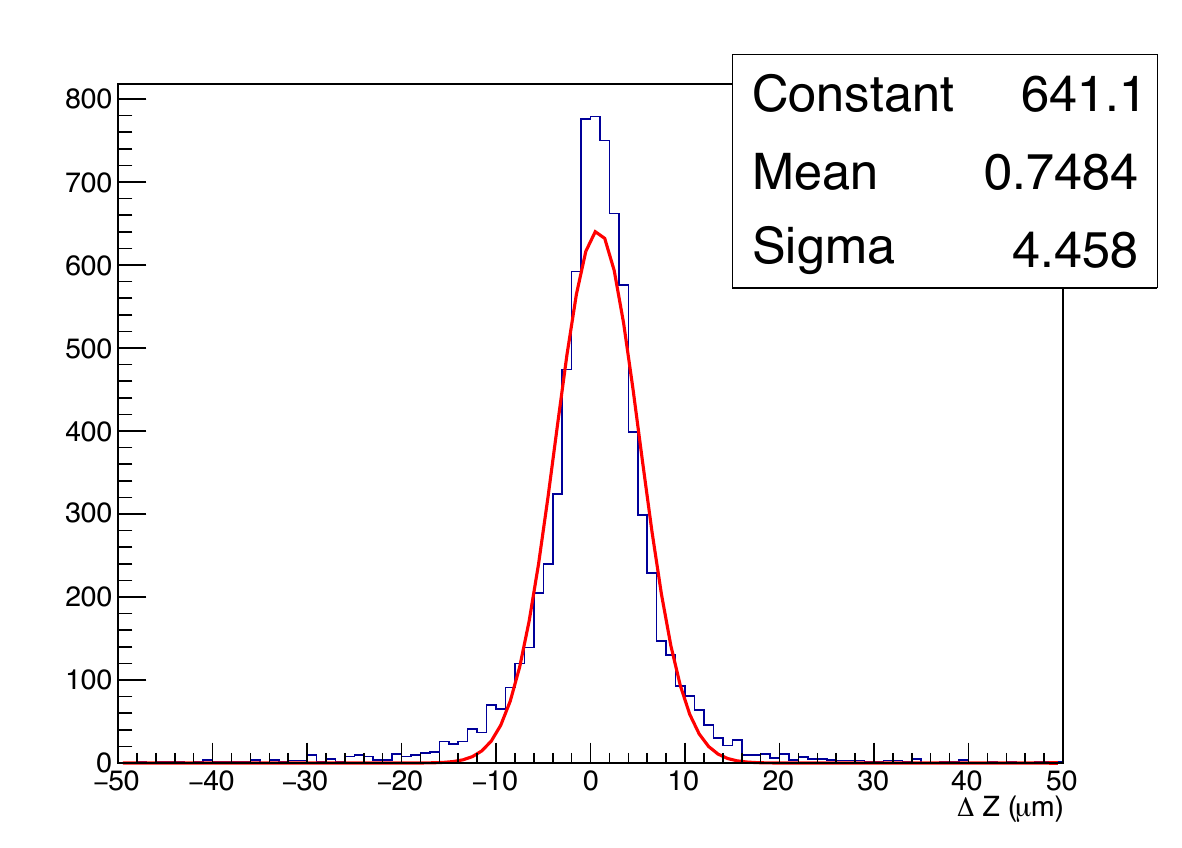}
    \caption{The difference between the true and reconstructed vertex positions in X, Y, and Z (listed from top to bottom)
is obtained using the EPOS MC. The mean and standard deviation values obtained from the Gaussian fit are used to
estimate the vertexing efficiency}
    \label{fig:4} 
\end{figure}

\begin{figure}[ht]
  \includegraphics[width=0.45\textwidth]{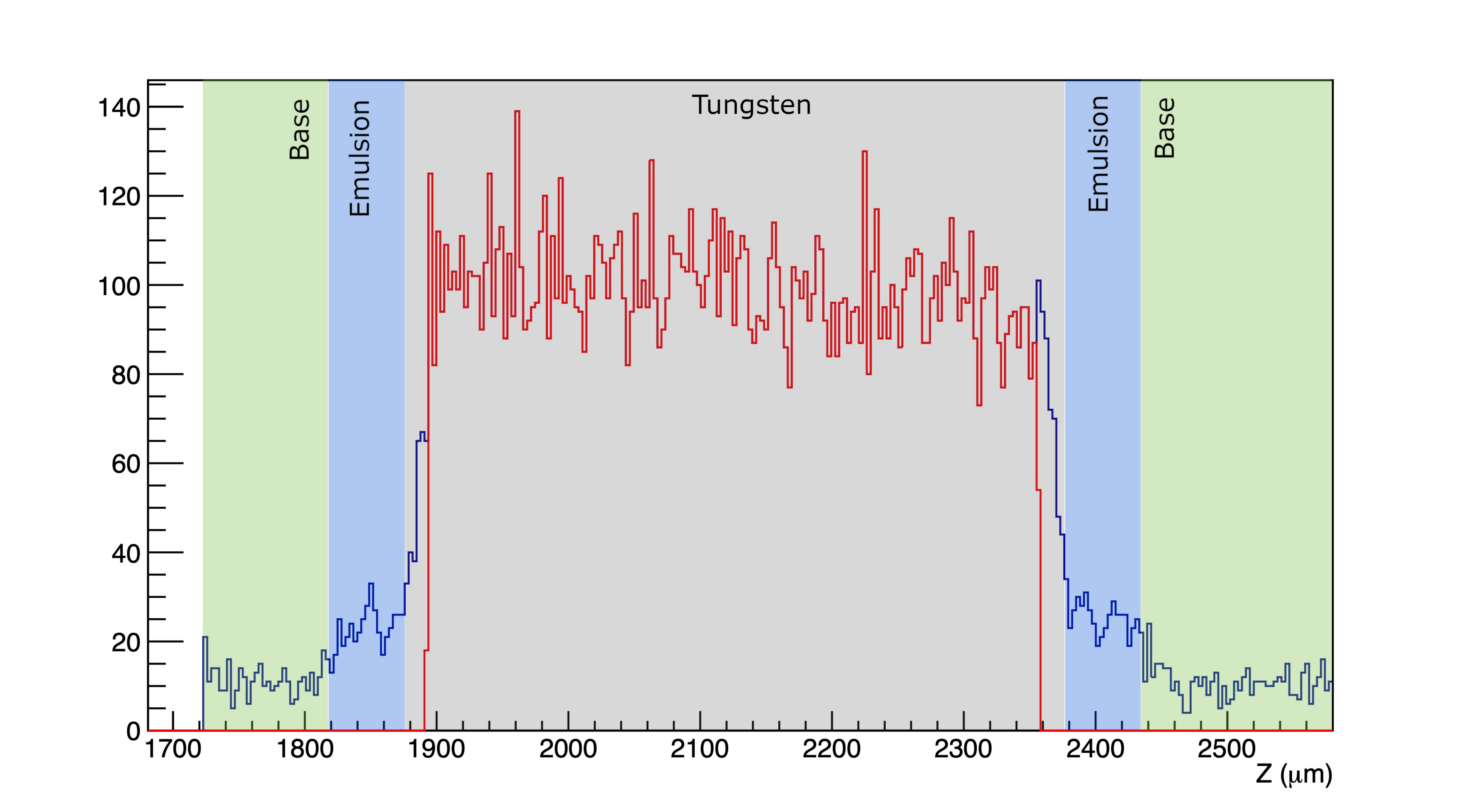}
\caption{The measured vertex positions in a tungsten plate and the adjacent emulsion plates along the longitudinal direction
in the data. Vertices within 4-sigma of vertex position resolution (red line), vertices outside the 4-sigma range (blue  line)}
\label{fig:7}       
\end{figure}

The efficiency of vertex reconstruction and proton- linking is evaluated  through a detailed simulation of the detector response using a program based on the GEANT4.11 toolkit~\cite{geant4}. The simulated geometry is
set as for the 2018 pilot run setup. About 2$\times 10^6$ proton interactions are generated using EPOS \cite{epos},  PYTHIA8 \cite{pythia}, QGSJET \cite{qgsjet},  DPMJET \cite{dpmjet} and GEANT4.11 (using the particle gun and relevant hadronic physics lists: G4Hadron-ElasticPhysics, G4HadronPhysicsFTFP-BERT, and G4-IonPhysics)  generators, considering the realistic beam proton density in a module.  
 The generated output is then transported through the module with GEANT4.11, and HTS tracking  algorithms are utilized.
 Data-driven smearing in coordinates and direction of the base-tracks has been applied  to reproduce the effect of measurement accuracy.
The angular smearing ranges from 1 to 7 mrad, while the position smearing varies between 0.5 and 2 $\mu$m, depending on the track angle.
Subsequently, the reconstruction algorithm, identical to the one used for data, is applied to MC samples to reconstruct particle tracks and vertices.
The vertex position resolution is determined
by comparing the true vertex position with the reconstructed one using MC.
A vertex is considered correctly reconstructed if it is found within the 4-sigma of vertex resolutions in the transverse plane  and the  z-direction  of the true vertex position as shown in Fig.~\ref{fig:4}.  The position resolution leads to the mis-reconstruction of vertex positions; some proton interactions in the emulsion (tungsten) are inaccurately reconstructed within the tungsten (emulsion). Furthermore, the thickness of the tungsten plates varies from one plate to another. To mitigate  these effects  a cut of 18 $\mu$m (equivalent to 4-sigma of the vertex resolutions) is applied to the vertex positions on both sides of the tungsten plate, as demonstrated in Fig.~\ref{fig:7}.  
The efficiency of this geometrical criteria is evaluated to be 92.8$\pm 0.2\%$.
Furthermore, a final correction for the coordinate system of the tungsten plate must be made due to the difference in refractive index between the emulsion and the plastic base. The coordinate system along the beam direction is multiplied by a factor of  1.052 $\pm$  0.002 \cite{GRAINE}. 
After applying this correction, the effective thickness of the tungsten plate is determined to be 489 $\pm$1 $\mu$m. For MC events, this correction is not applied, and the thickness is taken to be 464 $\mu$m.
\begin{table}[ht]
\caption{The number of reconstructed beam-proton interaction vertices (N) in the data for each tungsten plate, which
is 489$\pm$1 $\mu m$-thick, along with the number of beam protons ($N_0$) entering the tungsten plates of the module and the ratio($\frac{N}{N_0}$)}
\begin{tabular}{llll}
\hline\noalign{\smallskip}
Tungsten    & ~ & ~     & ~                     \\
Plate & N & $N_0$ & $\frac{N}{N_0}$ ($\%$)\\
\noalign{\smallskip}\hline\noalign{\smallskip}
1 & 13,586 & 3,310,658 & 0.41\\
2 & 13,390 & 3,292,677 & 0.41\\
3 & 12,653 & 3,256,746 & 0.39\\
4 & 12,256 & 3,214,141 & 0.38\\
5 & 11,745 & 3,157,020 & 0.37\\
6 & 11,264 & 3,082,105 & 0.36\\
7 & 10,645 & 2,996,099 & 0.35\\
8 & 9,775 & 2,892,348 & 0.34\\
\hline
Total & 95,314 & 25,201,794 & 0.38 \\
\noalign{\smallskip}\hline
\end{tabular}
\label{tab:11} 
\end{table}

\begin{figure}[ht]
  \includegraphics[width=0.50\textwidth]{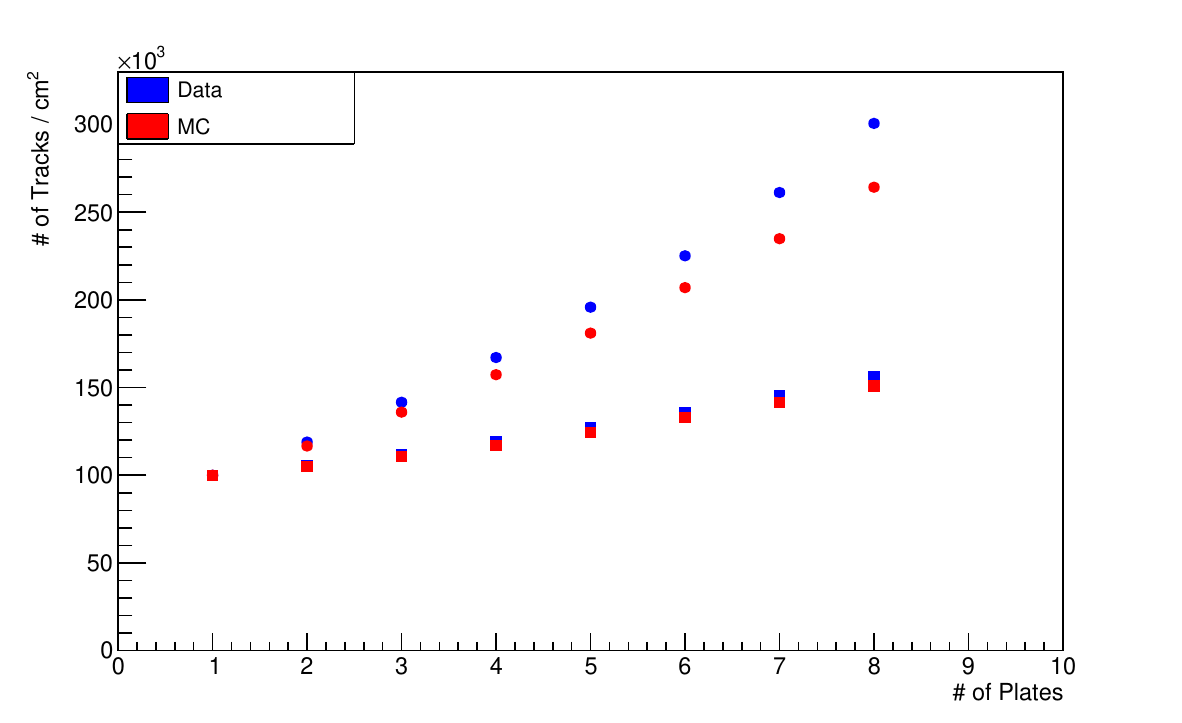}
\caption{Density of reconstructed tracks as a function of passing through tungsten plates of the 2018 run.
Solid circles denote  the density in all angular space (0$<$tan$\theta<$0.4), while solid squares represent  density of beam protons 
(with  tan$\theta<$0.03). The track densities are normalized to $10^5$ tracks/$\mathrm{cm^2}$ at the first tungsten plate}
\label{fig:8}       
\end{figure}
\begin{figure}[ht]
  \includegraphics[width=0.45\textwidth]{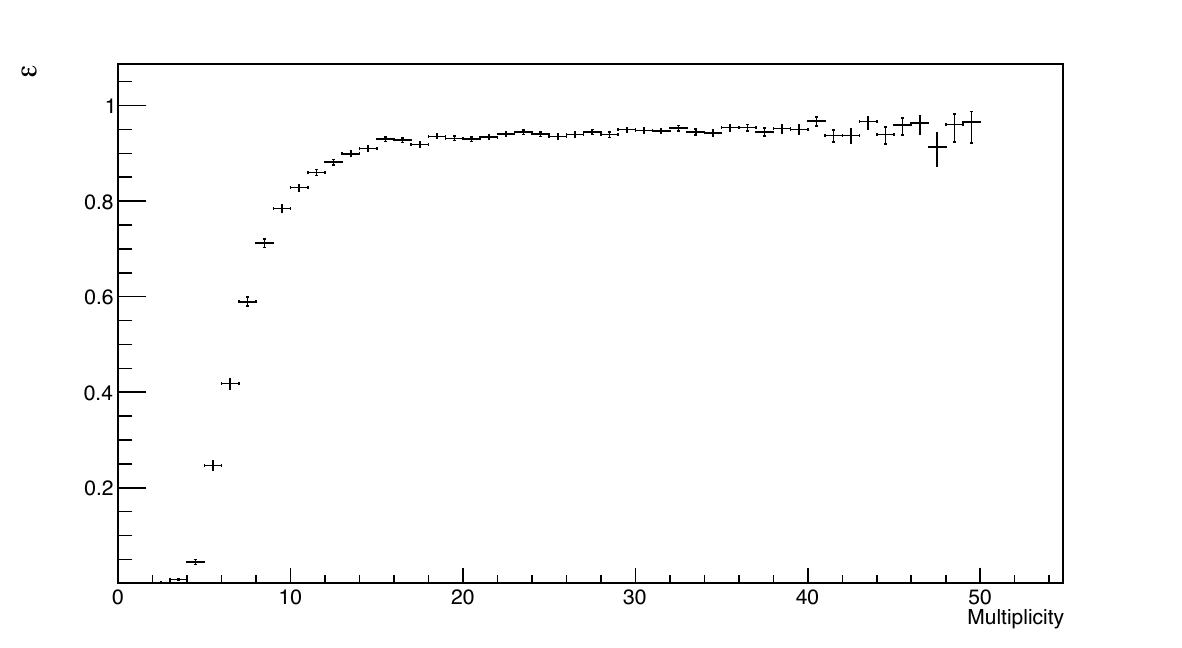}
\caption{Vertex reconstruction efficiency  as a function of  the reconstructed multiplicity of charged tracks at the vertex, obtained with EPOS MC}
\label{fig:9}       
\end{figure}

The efficiency estimation was performed using the EPOS event generator, which gives the best representation of the data distributions discussed in Section 3.2 
among the five event generators evaluated. From MC simulation, it was observed that the efficiencies depend on the track density in the module. Although the track density in the MC simulation does not fully represent the density observed in the data across all angular spaces, as shown in Fig.~\ref{fig:8} (solid circles), there is a good agreement in track density within the angular space of the beam protons  as indicated by the solid squares in Fig.~\ref{fig:8}. 
There is an excess of large-angle tracks, possibly resulting from secondary interactions or electron-positron pair production, which are not accurately modeled in the MC generator.
The efficiencies of proton linking and proton-vertex linking are primarily influenced by the track density of beam protons (tan$\theta <$0.030). In contrast, vertex reconstruction is affected by the density of all tracks (0$<$tan$\theta <$0.4) within the volume. Consequently, the track density of beam protons is employed to estimate linking efficiencies, while the overall track density across the entire angular space is used to evaluate vertex reconstruction efficiency.
The efficiencies are parameterized as a linear function of track density, and the actual track density from the data is subsequently used to estimate the efficiency for each tungsten plate. The results of these estimations are presented in Table~\ref{tab:12}. The proton-linking  efficiency is  more than $90 \% $  and almost constant  along the longitudinal direction.  The vertex reconstruction efficiency, which is approximately 80$\%$, depends on the multiplicity of the charged particles as shown Fig.~\ref{fig:9}. It increases with multiplicity up to 12, after which it reaches a plateau at approximately 95$\%$. Since a minimum of five tracks is required for the vertex reconstruction, the hadronization modeling implemented in the event generator may effect multiplicity distribution and, consequently, the efficiency of vertex reconstruction. The systematic error associated with this bias was calculated 2.5$\%$ by comparing the fraction of events with a multiplicity of greater than four among EPOS, DPMJET, and PYTHIA. 

\begin{table}[ht]
\caption{Vertex reconstruction($\epsilon_v$), Proton-Linking ($\epsilon_p$) and Proton-Vertex linking($\epsilon_{pv}$) efficiencies calculated using EPOS and realistic track density($\rho$) in each tungsten plate of a module}
\begin{tabular}{llll}
\hline\noalign{\smallskip}
Tungsten &  ~                  & ~                  & ~                  \\
Plate    & $\epsilon_v$ ($\%$) &$\epsilon_p$ ($\%$) & $\epsilon_{pv}$ ($\%$)\\
\noalign{\smallskip}\hline\noalign{\smallskip}
1 & 81.8 $\pm$ 2.5 & 93.7 $\pm$ 0.1 & 100.0 $\pm$ 0.1\\
2 & 81.4 $\pm$ 2.5 & 93.5 $\pm$ 0.1 & 100.0 $\pm$ 0.1\\
3 & 81.0 $\pm$ 2.5 & 93.2 $\pm$ 0.1 & 99.2 $\pm$ 0.1\\
4 & 80.6 $\pm$ 2.5 & 92.9 $\pm$ 0.1 & 98.2 $\pm$ 0.1\\
5 & 80.1 $\pm$ 2.5 & 92.6 $\pm$ 0.1 & 96.9 $\pm$ 0.1\\
6 & 79.6 $\pm$ 2.5 & 92.3 $\pm$ 0.1 & 95.7 $\pm$ 0.1\\
7 & 79.0 $\pm$ 2.5 & 92.0 $\pm$ 0.1 & 94.1 $\pm$ 0.1\\
8 & 78.4 $\pm$ 2.5 & 91.6 $\pm$ 0.1 & 92.2 $\pm$ 0.1\\
\hline
Mean & 80.3 $\pm$ 2.5 & 92.7 $\pm$ 0.1 & 97.0 $\pm$ 0.1\\
\noalign{\smallskip}\hline
\end{tabular}
\label{tab:12} 
\end{table}
\subsection{Data-MC comparisons}

\begin{figure}[ht]
  \includegraphics[width=0.45\textwidth]{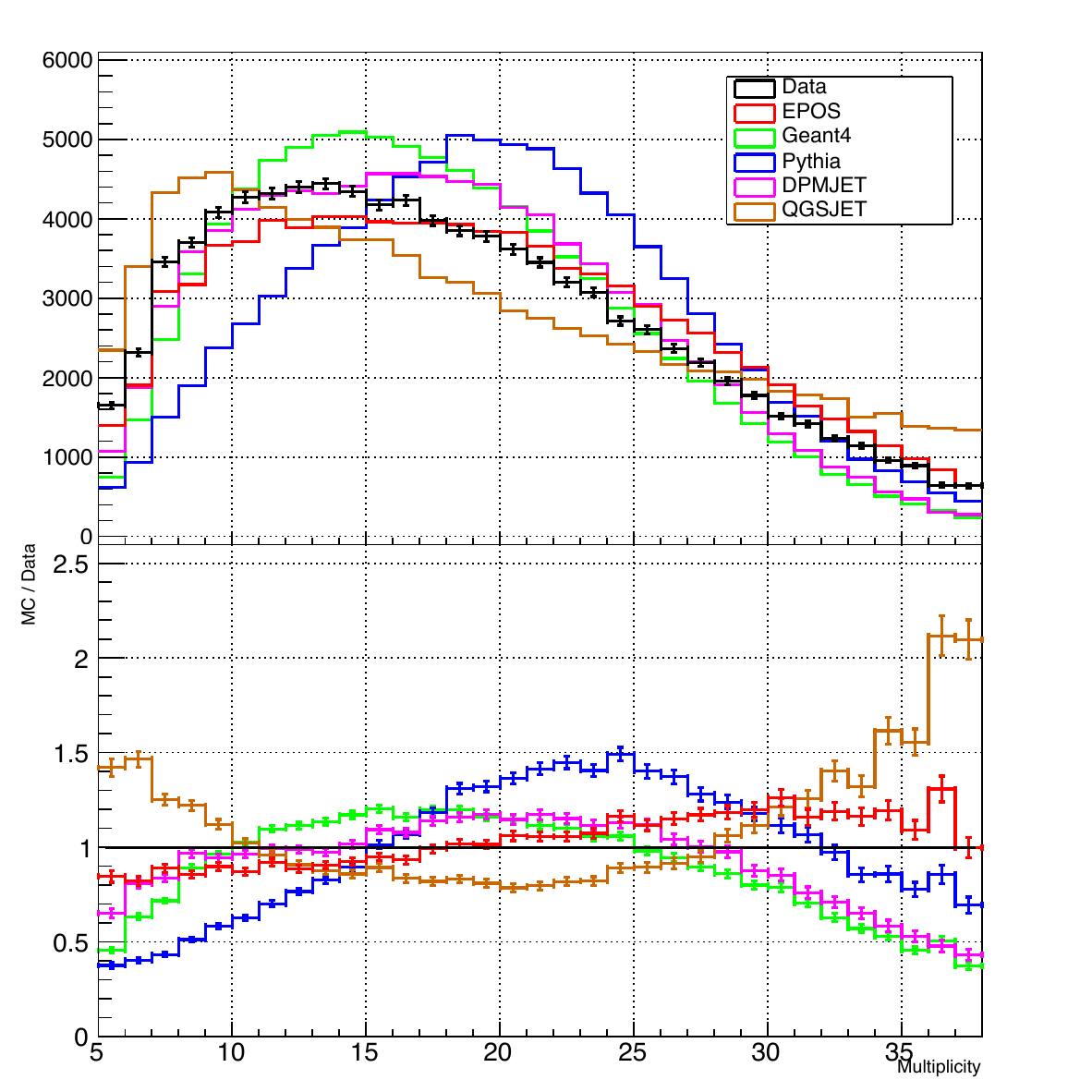}
\caption{MC/Data comparison of track multiplicity, normalized to the number of events in data}
\label{fig:10}       
\end{figure}

\begin{figure}[ht]
  \includegraphics[width=0.45\textwidth]{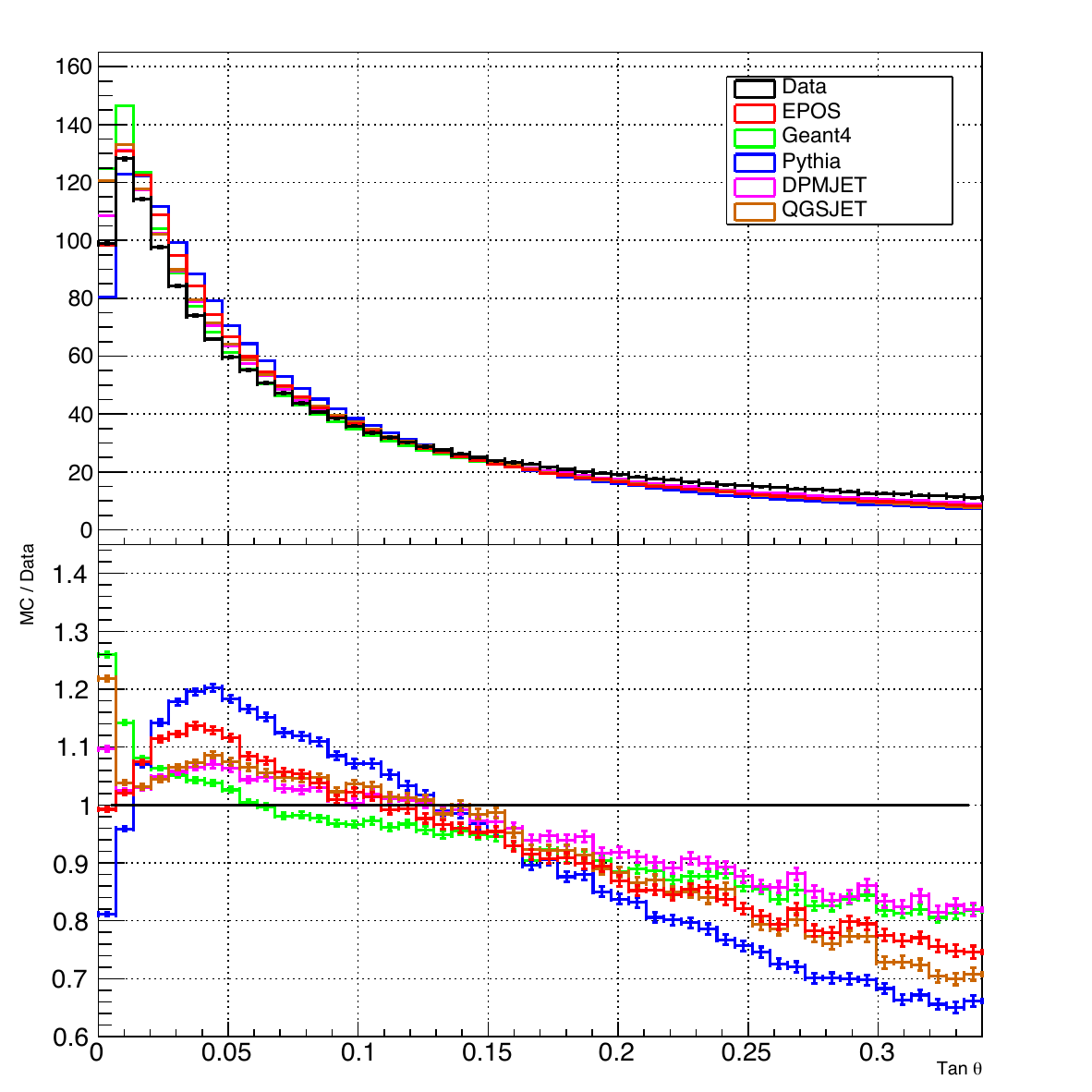}
\caption{MC/Data comparison of angular distribution of interaction daughters, normalized to the number of tracks in data}
\label{fig:11} 
\end{figure}

\begin{figure}[ht]
\includegraphics[width=0.45\textwidth]{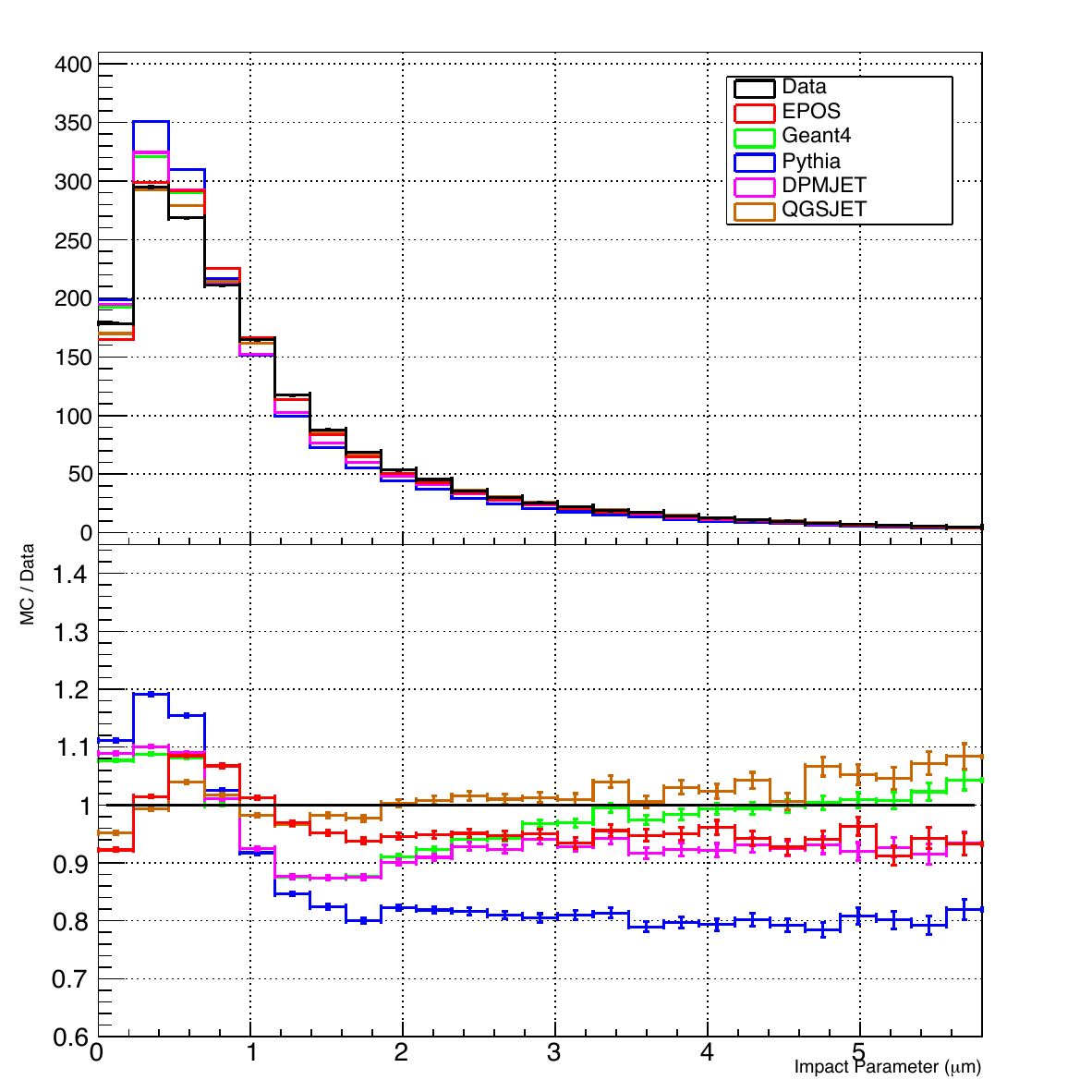}
\caption{MC/Data comparison of impact parameter of the particle tracks to interaction vertex, normalized to the number of tracks in data}
\label{fig:12}       
\end{figure}

\begin{figure}[ht]
  \includegraphics[width=0.45\textwidth]{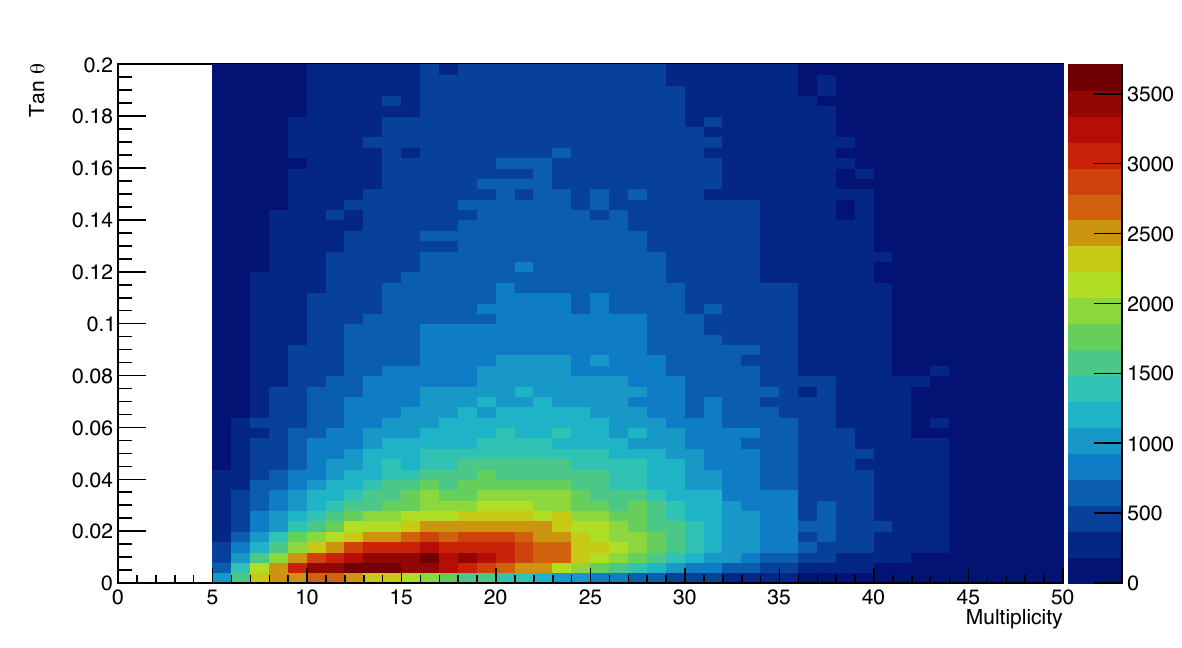}
\caption{Track slope vs multiplicity distribution in Data}
\label{fig:13}       
\end{figure}

\begin{figure}[ht]
  \includegraphics[width=0.45\textwidth]{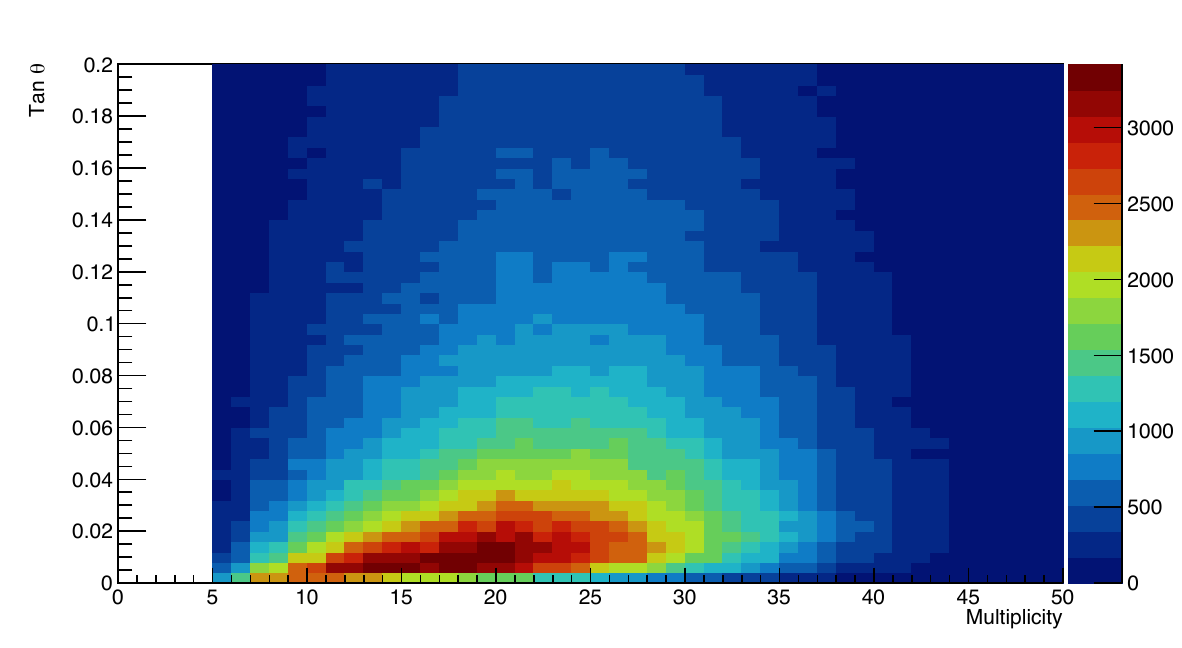}
\caption{Track  slope vs multiplicity distribution in EPOS}
\label{fig:14}       
\end{figure}

\begin{figure}[ht]
  \includegraphics[width=0.45\textwidth]{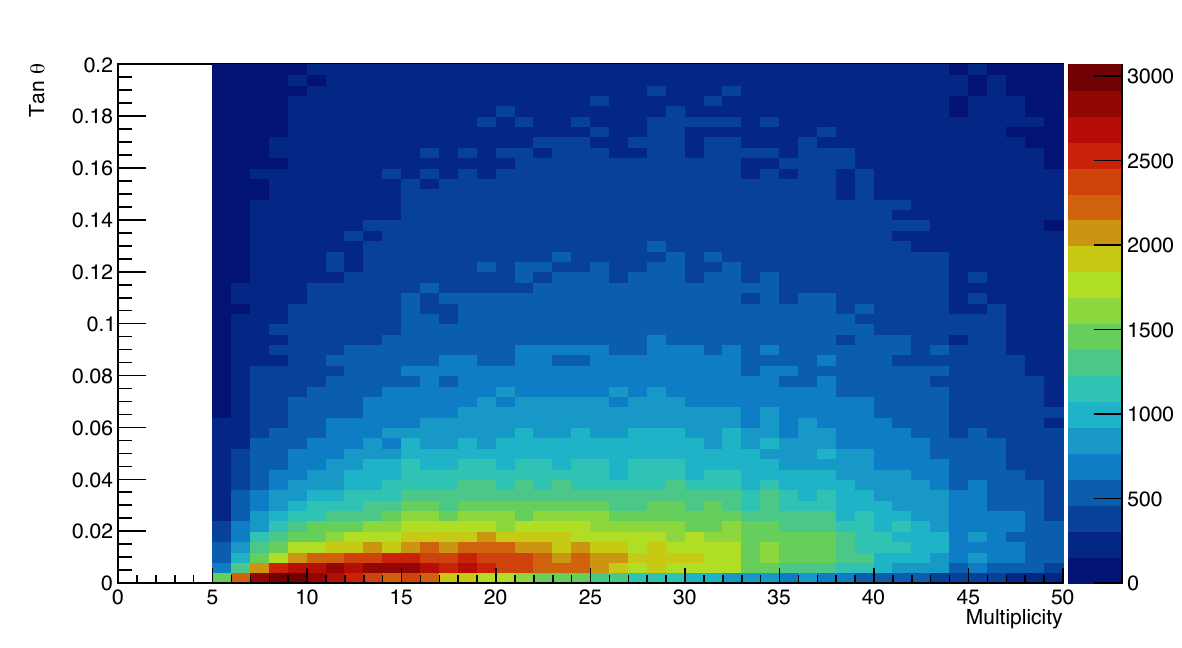}
\caption{Track  slope vs multiplicity distribution in QGSJET}
\label{fig:15}       
\end{figure}

\begin{figure}[ht]
\centering
\includegraphics[width=0.45\textwidth]{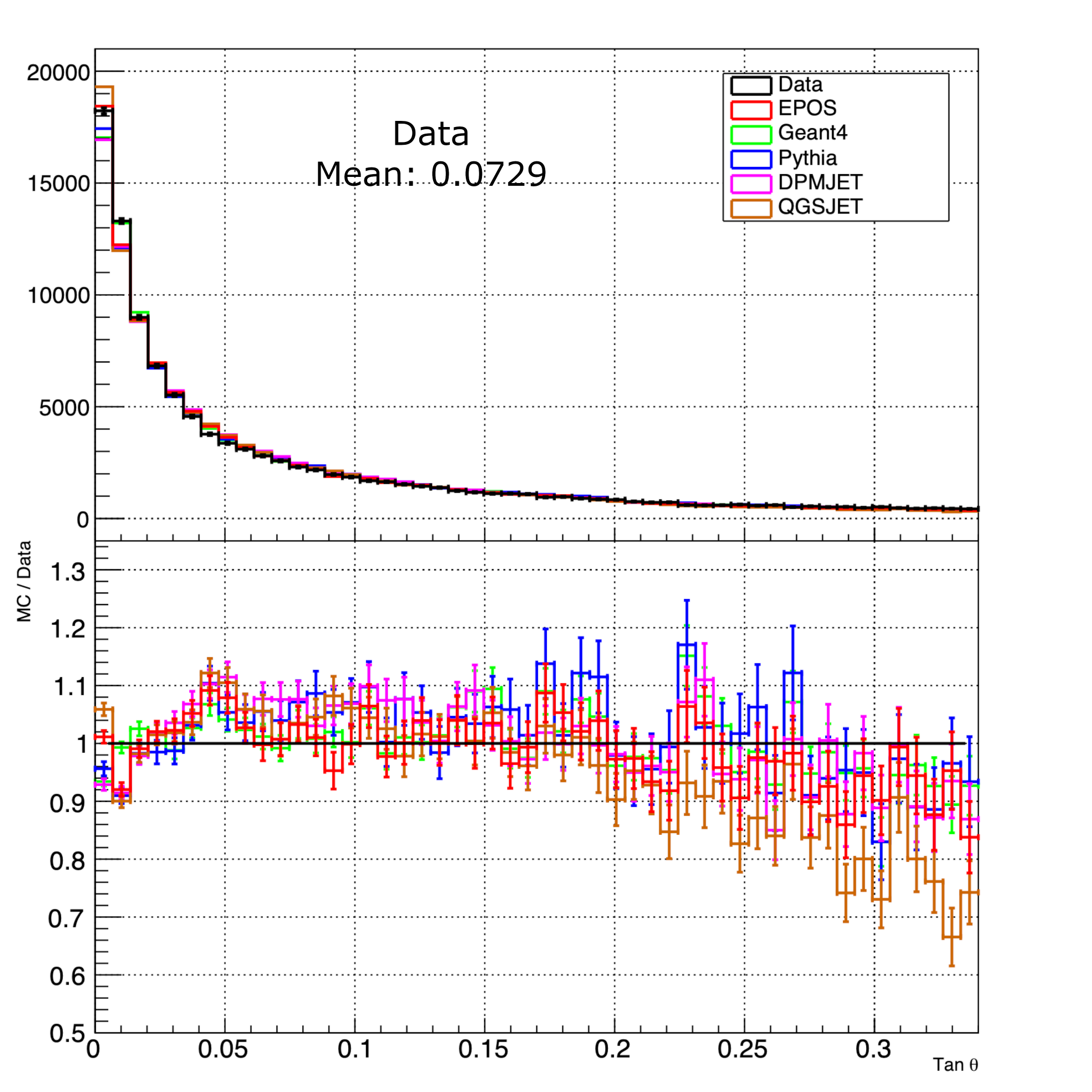}
\caption{MC/Data comparison of track angle for multiplicity less than 10}
\label{fig:16} 
\includegraphics[width=0.45\textwidth]{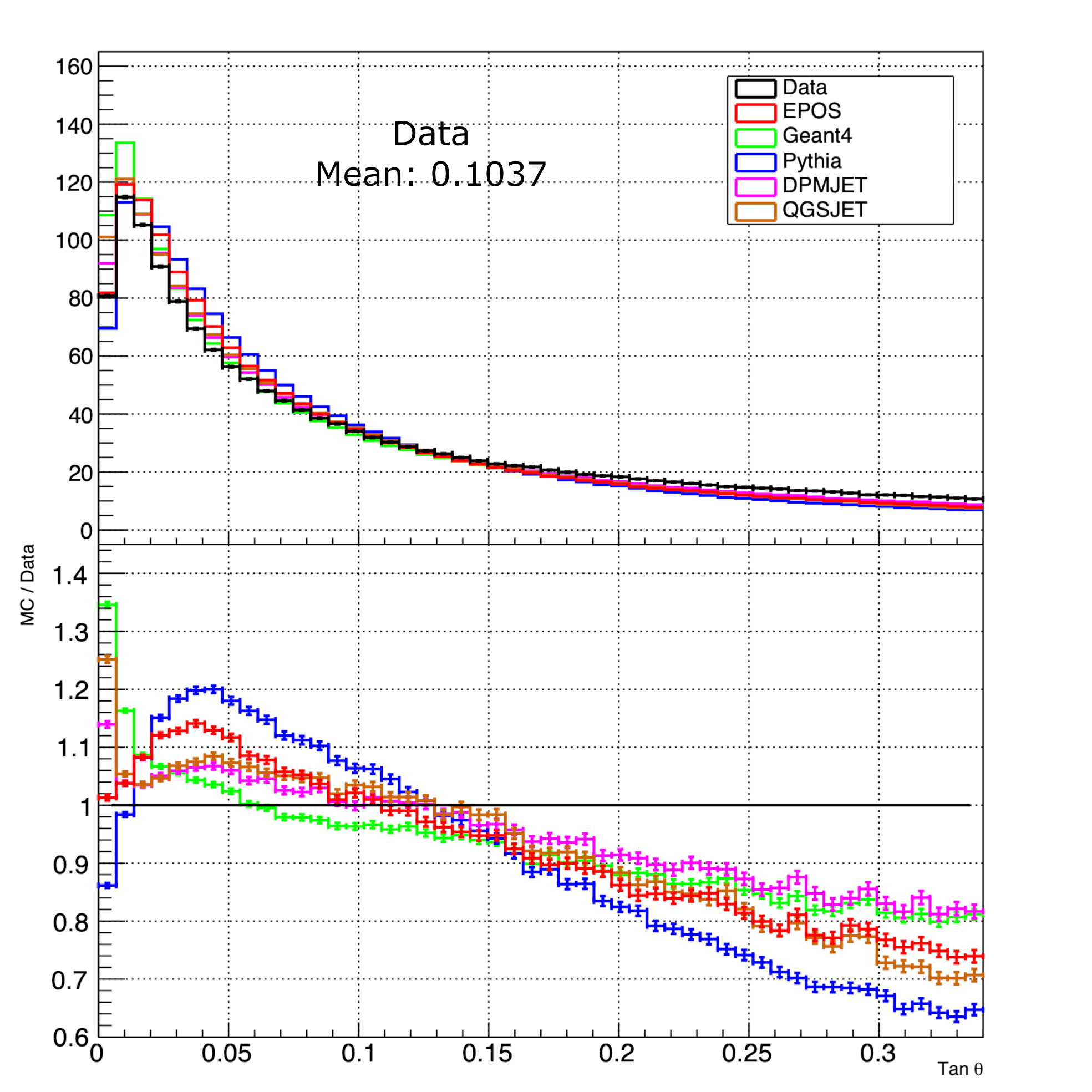}
\caption{MC/Data comparison of track slope for multiplicity greater than or equal to 10}
\label{fig:17} 
\end{figure}

The experimental data are compared with MC predictions in terms of multiplicity, particle angle,  and  impact parameter, as shown in Figs. ~\ref{fig:10}, ~\ref{fig:11}, and ~\ref{fig:12}, respectively. Among the event generators, EPOS shows the closest overall agreement with the experimental data, except for the impact parameter, where QGSJET provides the best agreement. However, discrepancies are observed in track angles between the experimental data and all MC generator predictions.  This discrepancy is further investigated to determine whether it arises from inefficiencies in track reconstruction or from underlying physics processes in the MC event generators. Figs. ~\ref{fig:13}, ~\ref{fig:14}, and ~\ref{fig:15} show the particle slope versus track multiplicity distribution for both data and MC predictions (EPOS and QGSJET). The data and EPOS predictions exhibit a similar trend of increasing charged particle angles with  multiplicity. This trend can be attributed to the constant transverse momentum ($P_T$) characteristic of hadron interactions, where the average transverse momentum  of a jet remains nearly constant during the hadronization process \cite{constPt}. In contrast, the QGSJET predictions do not follow the same trend as the data and EPOS results. To further investigate the discrepancies between the data and MC predictions in track angles, the data sample is divided into low-multiplicity and high-multiplicity events. 
Figs. ~\ref{fig:16} and ~\ref{fig:17} compare the angles of secondary tracks between the data and MC for vertices with track multiplicities less than 10 and 10 or more, respectively. The agreement between the data and MC simulations is good for all event generators in low multiplicity vertices.
However, in high multiplicity events, the agreement between MC  and data deteriorates, although it remains satisfactory up to 200 mrad, with the exception of QGSJET.


\subsection{The validity of KNO-G scaling}
\begin{figure}[ht]
  \includegraphics[width=0.45\textwidth]{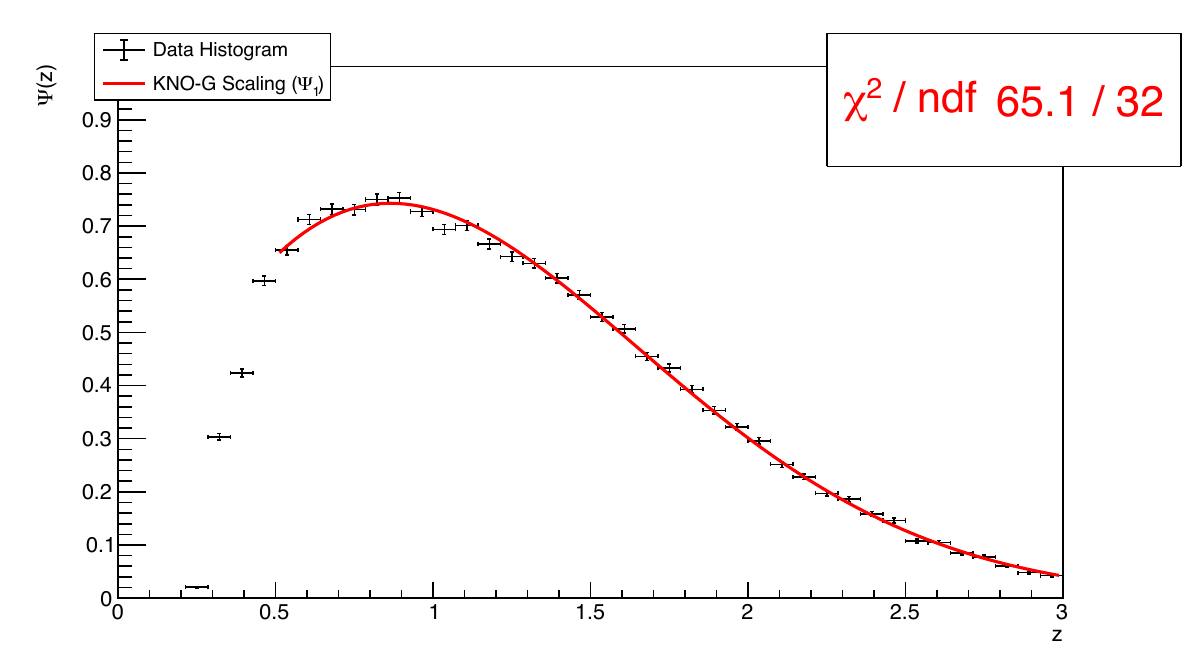}
\caption{KNO-G fits on the scaled multiplicity}
\label{fig:18}       
\end{figure}

The scaling behavior provides insights into the underlying dynamics of particle production and can help constrain theoretical models and understand the nature of particle interactions at high energies. Koba-Nielsen-Olesen (KNO) scaling~\cite{kno}  was formulated for the asymptotic energies,  at a finite energy range, however, its formulation is not self-consistent.  
Later,  it was reformulated   by Golokhvastov
to make it  self-consistent at all energies\cite{knog}. 
In the past, the validity of the KNO and KNO-G scalings was  tested in various high-energy experiments with different beams and energies ~\cite{knoexp1,knoexp2,knoexp3,knoexp4,slater,knog2}. In general, KNO-G scaling has been found to hold with reasonable accuracy, typically within a range of about 5-10$\%$.
We have also analyzed our data  to look for their  scaling behavior.   
The KNO-G scaling predictions are  tested by fitting  the  following scaling function~\cite{knog2}, which depends  on the scaled multiplicity 
z = $\frac{n_s}{<n_s>}$   ($n_s$ is the multiplicity of charged particles).
    \begin{equation}
      \Psi (z)= a_{1}z^{a_{3}}e^{-a_{2}z^2} 
\end{equation}
The mean multiplicity $<n_s>$ was calculated taken into account the effect of $n_s<5$. 
The correction factor was estimated as 0.95$\pm$0.02 by comparing mean values of multiplicity distributions  from true MC with and without $n_s <5$ selection.
The fit was done over  the range 0.5 $<$ z $<$ 3.0 where  the multiplicity is measured with  high efficiency, as shown in Fig.~\ref{fig:18}.
The fit values are presented in Table \ref{tab:1} for comparison with the values in~\cite{knog2}, which did not report errors in the fit parameters.
The fit has a $\chi^{2}$ value of 65.1 with 32 degrees of freedom.
Our multiplicity distribution is  found to be  nearly consistent with the KNO-G  scaling predictions.
\begin{table}[ht]
\caption{The obtained KNO-G fit parameters and those reported in~\cite{knog2}}
\label{tab:1}       
\centering
\begin{tabular}{lll}
\hline
$\Psi$ & Our fit & \cite{knog2}\\
\hline
$a_{1}$ & 1.15$\pm$0.01   & 1.19\\
$a_{2}$ & 0.45$\pm$0.01   & 0.62\\
$a_{3}$ & 0.67$\pm$0.02   & 0.66\\
\hline
\end{tabular}
\end{table}
\subsection{Measurement of proton interaction length}

To determine the interaction length of protons in tungsten, the number of reconstructed beam protons entering the tungsten plate and their interaction vertices  in tungsten were used. For each tungsten plate, interaction length is calculated using 
\begin{equation}
    \lambda = -\frac{L}{ln(1-\frac{N^\prime}{N_{0}^\prime})}
\end{equation}
where L is the measured thickness of the tungsten plate,  $N^\prime$ and $N_{0}^\prime$ are the number of proton interactions in the tungsten plate, and the number of beam protons entering the tungsten plate  which are corrected by efficiencies of Table \ref{tab:12}. 
The value of  L is 489$\pm$1 $\mu$m (464 $\mu$m) for Data (MC).  
The measured mean interaction length in tungsten is 93.7 $\pm$ 2.6 mm, which is in good agreement with the EPOS prediction of 95.8 $\pm$ 2.8 mm. The interaction lengths across all tungsten plates for both the data and EPOS are nearly constant, as presented in Table ~\ref{tab:6}.
\begin{table}[ht]
\caption{Estimated proton interaction length in tungsten for Data and EPOS. The statistical uncertainty is negligible  when compared to the systematic error which is caused by uncertainties in the vertexing efficiency}
\label{tab:6}       
\begin{tabular}{lll}
\hline\noalign{\smallskip}
Sub-volume & Data(mm) & EPOS(mm)\\
\noalign{\smallskip}\hline\noalign{\smallskip}
1 & 91.0 $\pm$ 2.5 & 95.2 $\pm$ 2.7 \\
2 & 90.8 $\pm$ 2.5 & 95.5 $\pm$ 2.8 \\
3 & 93.7 $\pm$ 2.6 & 95.3 $\pm$ 2.8\\
4 & 93.9 $\pm$ 2.7 & 95.5 $\pm$ 2.8\\
5 & 94.5 $\pm$ 2.7 & 94.8 $\pm$ 2.8\\
6 & 94.4 $\pm$ 2.7 & 95.0 $\pm$ 2.8\\
7 & 94.7 $\pm$ 2.8 & 98.1 $\pm$ 3.1\\
8 & 96.8 $\pm$ 3.0 & 97.0 $\pm$ 3.1\\
\hline
Mean & 93.7 $\pm$ 2.6 & 95.8 $\pm$ 2.8\\
\noalign{\smallskip}\hline
\end{tabular}
\end{table}

\section{Conclusion}
The analysis results of the 2018  run demonstrate that precise tracking and vertexing can be successfully achieved even in high track density environments. This performance is essential for detecting charmed hadrons and accurately determining the tau neutrino flux.
Using a sub-sample of the 2018 run, we present first results on key distributions, along with comparisons between our data and the predictions from various event generators. In general, EPOS predictions align closely with our data across all distributions, with the exception of secondary particle slopes.
It is observed that the track angle shows a dependence on multiplicity in both data and MC. The agreement between MC and data in particle angle is within 10$\%$ for low multiplicity ($<$10) events, but it deteriorates to 20-30$\%$  for high multiplicity events. 
Our multiplicity distribution is tested for KNO-G scaling and is found to be nearly consistent. We also report the first measurement of the proton interaction length in tungsten, which is determined to be 93.7 $\pm$ 2.6 mm. The results presented in this study have been obtained primarily to assist in refining proton-nucleus interaction models used in MC event generators.

\begin{acknowledgements}

Funding is gratefully acknowledged from national agencies and Institutions supporting us,
namely: JSPS KAKENHI for Japan (Grant No. JP 20K23373, JP 18KK0085, JP 17H06926, JP
18H05541 and JP 23H00103), CERN-RO(CERN Research Programme) for Romania (Contract No. 03/03.01.2022)
and TENMAK for Turkey (Grant No. 2022TENMAK(CERN) A5.H3.F2-1).
\end{acknowledgements}





\end{document}